\documentclass{article}%
\usepackage{amsmath}
\usepackage{amsfonts}
\usepackage{amssymb}
\usepackage{graphicx}%
\providecommand{\U}[1]{\protect\rule{.1in}{.1in}}

\begin{document}
\begin{titlepage}
\vspace{.3cm} \vspace{1cm}
\begin{center}
\baselineskip=16pt \centerline{\Large\bf  Inflation without Selfreproduction} \vspace{2truecm} \centerline{\large\bf
\ Viatcheslav Mukhanov\ \ } \vspace{.5truecm}
\emph{\centerline{Theoretical Physics, Ludwig Maxmillians University,Theresienstr. 37, 80333 Munich, Germany }}
\end{center}
\vspace{2cm}
\begin{center}
{\bf Abstract}
\end{center}
We find a rather unique extension of inflationary scenario which avoids selfreproduction
and thus resolves the problems of multiverse, predictability and initial conditions. In this theory the amplitude of
the cosmological perturbations is expressed entirely in terms of the total duration of inflation.
\end{titlepage}

\section{Introduction}

The recent CMB observations have unambiguously proven the theory of the
quantum origin of the universe structure. According to this theory the initial
quantum fluctuations were amplified in the very early universe and provided us
with the seeds for galaxies \cite{M1}, \cite{M2}. The simplest mechanism of
amplification of these quantum fluctuations is realized in inflationary
cosmology according to which the early universe went through the stage of
accelerated expansion. On the other hand, the quantum fluctuations amplified
during inflation also lead to selfreproduction and do not allow inflation to
end up once it was started \cite{Vil1}. Inflation continues forever leading to
a metaphysical (non-verifiable) concept of eternal universe and multiverse.
According to \cite{ILS1}, \cite{ILS2} the multiverse and eternal inflation
damage the predictive power of the theory because in this case
\textquotedblleft anything can happen and will happen an infinite number of
times\textquotedblright\ \cite{GKN}. Moreover, the most favored by recent
observations plateau-like potential returns us back the initial condition
problem:\ \textquotedblleft by favoring only plateau-like models, the
Planck2013 data creates a serious new challenge for the inflationary paradigm:
the universally accepted assumption about initial conditions no longer leads
to inflation; instead, inflation can only begin to smooth the universe if the
universe is unexpectedly smooth to begin with!\textquotedblright\ \cite{ILS1}.
Raising these issues the authors of \cite{ILS1}, \cite{ILS2} pose completely
legitimate question whether the eternal multiverse inflation, which they call
the \textquotedblleft postmodern inflationary paradigm\textquotedblright\ as
opposed to \textquotedblleft classic inflationary paradigm\textquotedblright,
does really allow us to explain anything and make any predictions? Surely one
could abandon this question restricting oneself only to the last hundred
e-folds of inflation responsible for the observable universe and refuse to
approximate the theory beyond the scales which we will never see anyway.
However, if we want to explain not only the quantum origin of the universe
structure, having no alternatives anyway, but also explain the origin of the
whole universe we have to understand whether inflation can in principle avoid
the problems raised and discussed in \cite{ILS1}, \cite{ILS2}, \cite{GKN}.

The purpose of this paper is to find a natural extension of inflation to the
Planck energy, which avoids the selfreproduction and initial condition problem
and thus allows us to turn back to the \textquotedblleft classic inflationary
paradigm\textquotedblright\ with all its predictive power.

\section{Inflation and selfreproduction}

Production of a closed universe does not cost any energy because the positive
energy of the matter is entirely compensated by the negative energy of the
gravitational self-interaction of this matter. Therefore the closed universe
can be produced as a result of quantum fluctuations \cite{T}. One can expect
that quantum fluctuations are essential only at Planck scale and only quantum
universes with internal mass of order $10^{-5}g$ can easily emerge. If gravity
is an attractive force then such universes will immediately recollapse.
However, as it was pointed out in \cite{BEG},\cite{zel},\cite{Vil2}, this does
not happen if for some reason the equation of state within the Planckian
universe corresponds to the cosmological constant $p\approx-\varepsilon,$
where $p$ is the pressure and $\varepsilon$ is the energy density. In this
case the gravitational field, determined by $\varepsilon+3p\approx
-2\varepsilon,$ is repulsive and instead of collapsing the universe starts to
expand with acceleration. As a result the size of the closed universe grows
exponentially and its total mass (as well as the number degrees of freedom)
also increases exponentially fast. The energy needed to produce the matter
comes from the gravitational reservoir with unbounded from below energy. This
is a rough picture of the emergence of the causal universe in Minkowski space
\cite{BEG} or from nothing \cite{zel},\cite{Vil2}\footnote{The idea of the
creation of universe from nothing was suggested independently by L. Grishchuk,
Y. Zeldovich \cite{zel} and A. Vilenkin\cite{Vil2}, who also offered the
mathematical description for the nucleation process.}, which was proposed in
80th. This picture is also well supported by quantization of space-time in
noncommutative geometry \cite{CCM}. The stage of accelerated expansion is
useful for amplifying quantum fluctuations which later serve as the seeds for
galaxy formation \cite{M1}, \cite{M2} and, moreover, it also amplifies the
quantum fluctuations of transverse degrees of freedom of the gravitational
field (gravitational waves) \cite{Star}.

There are many inflationary scenarios in the literature the only purpose of
which is to provide us with the stage of the quasi-exponential expansion.
These scenarios mainly differ by the choice of a slow-roll scalar field
potential \textquotedblleft justified\textquotedblright\ by a
\textquotedblleft fundamental theory\textquotedblright. However such theory is
not yet known and hence any particular potential is merely based on the
prejudices of the authors. I believe that under such circumstances the more
plausible approach is an effective description of inflation in terms of the
effective equation of state parametrized by the number of e-folds left until
the end of inflation. As it was shown in \cite{Muk0} even the simplest choice
for the equation of state allows us to cover nearly all scenarios and to prove
that the most valuable predictions of the theory of quantum origin of the
universe structure are robust. From the very beginning of the accelerated
expansion (inflation) there must be a small deviation of the equation of state
from the cosmological constant, i.e.,
\begin{equation}
1+w\equiv1+\frac{p}{\varepsilon}\ll1,\label{1a}%
\end{equation}
\textit{but nonvanishing,} because otherwise inflation would never end. To
describe how this deviation is changing with time we will use as a time
parameter the number of e-folds $N$ left to the end of inflation, defined as
\begin{equation}
a=a_{f}\exp\left(  -N\right)  ,\label{1}%
\end{equation}
where $a$ is the scale factor and $a_{f}$ is its value at the end of inflation
when $1+w\simeq O\left(  1\right)  .$ The interval of $N$ relevant for the
observable universe is not very large, namely, $N<70.$ At the end of inflation
the squared amplitude of the Newtonian gravitational potential in scales
$\lambda$ is (see \cite{Mbook})%
\begin{equation}
\Phi_{\lambda}^{2}\simeq\left.  \frac{\varepsilon}{1+w}\right\vert
_{N_{\lambda}}\label{2a}%
\end{equation}
where we have used the Planck units in which $\varepsilon_{Pl}=1$ and
$N_{\lambda}$ is the number of e-folds left to the end of inflation since the
time when the commoving scale $\lambda$ crosses the Hubble scale
$H^{-1}=a/\dot{a},$ that is,%
\begin{equation}
H^{-1}=a_{f}e^{-N_{\lambda}}\lambda.\label{2aa}%
\end{equation}
For observed in the CMB experiments scales $N_{\lambda}\simeq50\div60$.
Assuming that the equation of state is monotonic and smooth and taking into
account that $1+w\simeq O\left(  1\right)  $ when inflation ends at
$N\simeq1,$ it is rather natural to approximate it for $N>1$ as
\begin{equation}
1+w\simeq\frac{\beta}{N^{\alpha}},\label{2}%
\end{equation}
where $\alpha$ and $\beta$ are both the positive constants of order unity.
\ Using the energy conservation equation%
\begin{equation}
\dot{\varepsilon}=-3H\left(  \varepsilon+p\right)  ,\label{3a}%
\end{equation}
rewritten as%
\begin{equation}
\frac{d\ln\varepsilon}{dN}=3\left(  1+w\right)  \simeq\frac{3\beta}{N^{\alpha
}},\label{4a}%
\end{equation}
we find%
\begin{equation}
\varepsilon\left(  N\right)  \simeq\left\{
\begin{array}
[c]{c}%
\varepsilon_{0}N^{3\beta},\text{
\ \ \ \ \ \ \ \ \ \ \ \ \ \ \ \ \ \ \ \ \ \ \ \ }\alpha=1,\\
\varepsilon_{0}\exp\left(  -\frac{3\beta}{\alpha-1}\frac{1}{N^{\alpha-1}%
}\right)  ,\text{ \ \ \ \ \ }\alpha\neq1.
\end{array}
\right.  \label{5a}%
\end{equation}
Correspondingly the amplitude of the scalar perturbations is
\begin{equation}
\Phi_{\lambda}^{2}\simeq\left\{
\begin{array}
[c]{c}%
\varepsilon_{0}N^{3\beta+1},\text{
\ \ \ \ \ \ \ \ \ \ \ \ \ \ \ \ \ \ \ \ \ \ \ \ \ }\alpha=1,\\
\varepsilon_{0}N^{\alpha}\exp\left(  -\frac{3\beta}{\alpha-1}\frac
{1}{N^{\alpha-1}}\right)  ,\text{ \ \ \ \ \ }\alpha\neq1.
\end{array}
\right.  \label{6a}%
\end{equation}
The spectral index $n_{s},$ defined in the observable range of scales via
$\Phi^{2}\propto\lambda^{1-n_{s}},$ is then equal to%
\begin{equation}
1-n_{s}=\frac{d\ln\Phi^{2}}{d\ln\lambda}=\frac{d\ln\Phi^{2}}{dN_{\lambda}%
}\simeq\left\{
\begin{array}
[c]{c}%
\frac{3\beta+1}{N_{\lambda}},\text{ \ \ \ \ \ \ }\alpha=1,\\
\frac{\alpha}{N_{\lambda}},\text{ \ \ \ \ \ \ \ \ \ }\alpha>1.
\end{array}
\right.  \label{6aa}%
\end{equation}
It was shown in \cite{Muk0} that in terms of the slow-roll scalar field
potentials the case $\alpha=1$ corresponds to the chaotic inflation \cite{L1}
with power-law potentials, $\alpha=2$ describes the plateau-like potentials
and finally $\alpha>2$ describe the scenarios with small scalar field, in
particular, the new inflation \cite{L2}. According to the recent Planck data
\cite{Planck}, $n_{s}=0.96\pm0.007.$ Therefore the cases $\alpha=1$ for
$3\beta=1$ and $\alpha=2$ are strongly favored by the observations. The energy
density at the end of inflation, $\varepsilon_{0}$ in (\ref{5a}), is
determined by requiring $\Phi_{\lambda}^{2}\simeq10^{-9}$ for $N_{\lambda
}\simeq50$ to fit the observations. For both cases above we have,
\begin{equation}
\varepsilon_{0}\simeq10^{-12},\label{8a}%
\end{equation}
in Planck units. The models with $\alpha=1,$ $3\beta=1$ and $\alpha=2$ are
indistinguishable from the point of view of the scalar perturbations because
they result in precisely the same spectrum of inhomogeneities. However, they
predict different ratios of tensor to scalar perturbations (see \cite{Mbook}%
):
\begin{equation}
r=24\left(  1+w\right)  =\frac{24\beta}{N^{\alpha}}.\label{9a}%
\end{equation}
In case $3\beta=1$ we have $r=0.16.$ So high value of $r$ is seriously
disfavored by the Planck data according to which $r<0.11$ at $95\%$ confidence
level \cite{Planck}. For $\alpha=2$ the amount of gravitational waves is about
$N_{\lambda}\simeq50$ times less and corresponds to $r=0.003$ in perfect
agreement with observations. Thus, the recent CMB Planck results seems
strongly favor plateau-like potentials and rule out at significant confidence
level both the chaotic and new inflation.

As we have seen above, given the spectral index $n_{s}=0.96$ one can fix the
energy scale at the end of inflation to be about $\varepsilon_{0}%
\simeq10^{-12}.$ In turn this determines the lower bound on the amount of the
gravitational waves. Hence, taking into account the results of the CMB
observations we conclude that $r=O\left(  1\right)  \times0.003$ characterizes
not only the most likely expected level of the gravitational waves but also it
is \textit{the lower bound} on the amount of the tensor perturbations produced
during inflation.

It follows from (\ref{6a}) that the amplitude of perturbations on the scales
which left the Hubble scale at%
\begin{equation}
N>N_{sr}\simeq\left\{
\begin{array}
[c]{c}%
\varepsilon_{0}^{-\frac{1}{3\beta+1}},\text{ \ \ \ \ \ \ }\alpha=1,\\
\varepsilon_{0}^{-\frac{1}{\alpha}},\text{ \ \ \ \ \ \ \ \ \ }\alpha>1.
\end{array}
\right.  \label{10a}%
\end{equation}
exceeds unity and the universe on these scales should be very inhomogeneous.
This happens because for $N>N_{sr}$ the universe is in the selfreproduction
regime in which quantum fluctuations dominate over the classical evolution and
the energy density in the patches of Hubble size instead of decreasing can
increase as a consequence of the quantum jumps. As a result, there emerges
exponentially growing volume, where inflation never ends and finally this
leads to eternal infinite multiverse dominated by inflating regions. In the
cases $\alpha=1,$ $3\beta=1$ and $\alpha=2$ the minimal number of e-folds
which leads to the eternal inflation is rather large, $N_{sr}\simeq10^{6},$
compared to the number of e-folds $N\simeq10^{2}$ needed to explain the
observable universe. One, of course, could doubt the applicability of the
theory so far beyond the observable scales. However, if inflation pretends
that it really solves the initial condition problem for the whole universe as
it was claimed in the \textquotedblleft classic inflationary
paradigm\textquotedblright, one has either to resolve the measure problem in
the multiverse (and it is not even clear whether this problem can be well
formulated) or find a way to avoid the multiverse and eternity, thus escaping
the problems of infinities and measures.

\section{Avoiding selfreproduction}

At first glance it looks like the necessary and sufficient condition for the
selfreproduction,
\begin{equation}
\left.  \frac{\varepsilon}{1+w}\right\vert _{N_{sr}}\simeq1, \label{11a}%
\end{equation}
is always satisfied for some $N_{sr}$ especially if inflation begins at
$\varepsilon\simeq1$ to avoid the initial condition\ problem. In fact, for the
models considered above $1+w$ decreases with $N$ while $\varepsilon\left(
N\right)  $ either grows ($\alpha=1$) or approaches a constant value
($\alpha>1).$ In this case equation (\ref{11a}) always has a solution at
$\varepsilon\left(  N_{sr}\right)  <1$ and the selfreproduction is
unavoidable. If we assume that for large $N$ the equation of state approaches
constant value, $1+w_{0}\neq0,$ then the energy density would grow with $N$
and at $\varepsilon_{sr}\simeq1+w_{0}\ll1$ we again inevitably enter the
selfreproduction regime. Therefore, the only way to avoid this regime is to
assume that at some large $N,$ the value of $1+w\left(  N\right)  $ begins to
increase with $N$. As it follows from (\ref{4a}) the energy density is then
equal to the Planck density $\varepsilon\simeq1$ for some finite $N_{m}$.
Thus, if we want to avoid the selfreproduction and do not face the problem of
initial conditions the required equation of state must simultaneously satisfy
the following requirements:

a) $1+w\left(  N\right)  \simeq1$ at $N\simeq1$ (to have graceful exit),

b) $1+w\left(  N\right)  \leq2/3$ at $N\simeq N_{m}$ (to solve initial
condition problem),

c) $1+w\left(  N\right)  \ll1$ for $1<N<N_{m}$ (inflation),

d) $1+w\left(  N\right)  >\varepsilon\left(  N\right)  $ for $1<N<N_{m}$ (no selfreproduction).

It is clear that if $1+w\ll1$ at $\varepsilon\simeq1$ then the
selfreproduction would take place already at the energy density $\varepsilon
\simeq1+w\ll1.$ Therefore, we assume that at Planck scale $1+w\left(
N_{m}\right)  $ is for instance $1/3$, just to begin the accelerated expansion
and $1+w\ll1$ after that. The conditions b)-c) can simultaneously be satisfied
if we assume that
\begin{equation}
1+w\simeq\varepsilon^{\gamma},\label{13a}%
\end{equation}
with $\gamma<1.$ In fact, in this case $1+w\ll1$ for $\varepsilon\ll1$ and
hence inflation takes place; moreover,
\begin{equation}
\frac{\varepsilon}{1+w}\simeq\varepsilon^{1-\gamma}<1\label{14a}%
\end{equation}
for $\varepsilon<1,$ and \textquotedblleft no
selfreproduction\textquotedblright\ condition d) is also satisfied.
Substituting $\left(  \ref{13a}\right)  $ in $\left(  \ref{4a}\right)  $ and
integrating we find%
\begin{equation}
\varepsilon\left(  N\right)  \simeq\left(  \frac{1}{3\gamma\left(
N_{m}+1-N\right)  }\right)  ^{1/\gamma},\label{15a}%
\end{equation}
and
\begin{equation}
1+w\left(  N\right)  \simeq\frac{1}{3\gamma\left(  N_{m}+1-N\right)
}.\label{16a}%
\end{equation}
At the beginning of inflation, at $N=N_{m},$ both $\varepsilon$ and $1+w$ are
of order unity and for $N<N_{m}$ we have inflation with $1+w\simeq\left(
3\gamma N_{m}\right)  ^{-1}\ll1.$ One can wonder to what extent the choice
$\left(  \ref{13a}\right)  $ is ambiguous. Let us show that it is nearly
unique. With this purpose we consider more general function for $1+w\left(
N\right)  $ as, for example,%
\begin{equation}
1+w\left(  N\right)  \simeq\frac{a}{\left(  N_{m}+1-N\right)  ^{\delta}%
}\label{17a}%
\end{equation}
with $\delta\neq1.$ Then we find
\begin{equation}
\varepsilon\left(  N\right)  \simeq\exp\left(  \frac{3a}{\delta-1}\frac
{1}{\left(  N_{m}+1-N\right)  ^{\delta-1}}\right)  ,\label{18a}%
\end{equation}
where the constant of integration is fixed by imposing $\varepsilon\left(
N_{m}\right)  \simeq1$ to avoid the initial condition problem. It is clear
that for $\delta>1$ the energy density is of order unity for $N<N_{m},$ while
$1+w\ll1,$ and hence the selfreproduction is inevitable. For $\delta<1$ the
energy density drops too fast and for instance, for $\delta=1/2$ it is about
$\varepsilon\left(  N\right)  \simeq\exp\left(  -N_{m}^{1/2}\right)  $ at
$N\simeq N_{m}/2.$ Therefore the amplitude of perturbations generated at
inflation will be too small to explain the structure of the universe. In fact,
to be in agreement with observations $N_{m}^{1/2}$ must be larger \ than, at
least, hundred and hence $\varepsilon\simeq\exp\left(  -100\right)  $ is much
less compared to the required $10^{-12}.$ Thus, the behavior of the equation
of state given by $\left(  \ref{16a}\right)  $ is nearly unique at large $N.$
However, it does not satisfies the condition a) above. In fact, according to
$\left(  \ref{16a}\right)  $ $1+w$ approaches the constant for small $N$ and
inflation has no graceful exit. Therefore $\left(  \ref{16a}\right)  $ should
be modified for small $N.$ This can be easily achieved by combining $\left(
\ref{2}\right)  $ with $\left(  \ref{16a}\right)  $ and it is obvious that%
\begin{equation}
1+w\left(  N\right)  \simeq\frac{\beta}{N^{\alpha}}+\frac{1}{3\gamma\left(
N_{m}+1-N\right)  },\label{19a}%
\end{equation}
satisfies all conditions a)-d) simultaneously. In this case $1+w$ is of order
unity at the beginning of inflation at Planck density, then it decreases,
reaches the minimum and begins to grow towards the end of inflation and
finally becomes of order unity at $N\simeq1$. The energy density is given by
\begin{equation}
\varepsilon\left(  N\right)  \simeq\left\{
\begin{array}
[c]{c}%
\frac{1}{\left(  N_{m}+1-N\right)  ^{1/\gamma}}\left(  \frac{N}{N_{m}}\right)
^{3\beta},\text{ \ \ \ \ \ \ \ \ \ \ \ \ \ \ \ \ \ \ \ \ \ \ \ \ \ \ \ \ }%
\alpha=1,\\
\frac{1}{\left(  N_{m}+1-N\right)  ^{1/\gamma}}\exp\left(  -\frac{3\beta
}{\alpha-1}\frac{1}{N^{\alpha-1}}\right)  ,\text{
\ \ \ \ \ \ \ \ \ \ \ \ \ \ \ \ }\alpha\neq1,
\end{array}
\right.  \label{20a}%
\end{equation}
where the constant of integration is fixed by requiring $\varepsilon\left(
N_{m}\right)  \simeq1$.

In the case $\alpha=1$ and $3\beta=1,$ the amplitude of scalar perturbations
is about%
\begin{equation}
\Phi^{2}\simeq\frac{1}{\left(  N_{m}-N\right)  ^{\frac{1}{\gamma}-1}}\left(
\frac{N}{N_{m}}\right)  ^{2},\label{21a}%
\end{equation}
and correspondingly the spectral index is%
\begin{equation}
1-n_{s}\simeq\frac{2}{N}+\left(  \frac{1}{\gamma}-1\right)  \frac{1}{N_{m}%
-N}.\label{22a}%
\end{equation}
As we see in non-eternal inflating universe the amplitude of perturbations is
entirely determined by the duration of inflation and has the required value if%
\begin{equation}
N_{m}^{\frac{1}{\gamma}+1}\simeq10^{12}.\label{23a}%
\end{equation}
Because $\gamma\,<1$ (see $\left(  \ref{13a}\right)  $) the maximal duration
of inflation cannot exceed $10^{6}$ e-folds. For $1/\gamma=2$ the maximal
number of e-folds, which determines the size of the universe, is about
$10^{4}.$ Because the number of e-folds should definitely be larger that
$10^{2}$ we find that $1/\gamma$ must be smaller than $5.$ The model above
describes inflation which begins at Planck density (hence no initial condition
problem) and creates one unique universe (avoiding the selfreproduction and
multiverse problems). However, this model is disfavored by the CMB
observations because it produces too much gravitational waves. The most
favorable (nearly unique) model, which is in agreement with the observations,
corresponds to $\alpha=2.$ In this case
\begin{equation}
\Phi^{2}\simeq\frac{1}{\left(  N_{m}-N\right)  ^{\frac{1}{\gamma}-1}}%
\frac{N^{2}}{N_{m}+N^{2}},\label{24a}%
\end{equation}
and%
\begin{equation}
1-n_{s}\simeq\frac{2}{N}\left(  1+\frac{N^{2}}{N_{m}}\right)  ^{-1}+\left(
\frac{1}{\gamma}-1\right)  \frac{1}{N_{m}-N}.\label{25a}%
\end{equation}
Here the duration of inflation is determined by the condition%
\begin{equation}
N_{m}^{\frac{1}{\gamma}}\simeq10^{12},\label{25}%
\end{equation}
and hence for $1/\gamma=2$ the maximal number of e-folds is $10^{6}$ which
would be about the scale of selfreproduction if $1+w$ is given by $\left(
\ref{2}\right)  .$ However, in our case the second term in $\left(
\ref{19a}\right)  $ is larger than the first one for $N>N_{m}^{1/2}%
\simeq10^{3}$ and hence it determines the evolution for $N>10^{3}.$ For
$1/\gamma=4,$ we obtain $N_{m}\simeq10^{3}$ and the spectral index would be
substantially corrected compared to the case $\left(  \ref{2}\right)  $ in
observable scales at $N\simeq50.$ Thus, we conclude that $1/\gamma$ must be
$2$ or $3.$

\section{Slow-roll potentials}

In the modern literature the \textquotedblleft playground of inflationary
scenarios\textquotedblright\ mainly consists of the slow-roll potentials which
sometimes are rather complicated and based on ugly constructions. The only
purpose of these constructions is to justify the quasi-exponential expansion.
Moreover, the only relevant for comparison with observations features of the
potential are its value and its first and second derivatives basically
\textit{at one point}. Thus forgetting the questionable justification of the
scenarios from the point of view of a \textquotedblleft fundamental
theory\textquotedblright, the scalar field potentials are not more than merely
the way to parametrize the missing knowledge about fundamental physics at high
energies, not better than parametrization above. I believe that the scalar
field parametrization is not so natural as it looks at the first glance and
not even so convenient as used above. Therefore, perhaps it is not surprising
that all original scenarios which look simple in terms of the scalar field
have failed to describe the observations. In fact, everything we need for
explaining the observations is the description of inflation in terms of some
effective equation of state, because gravity seems entirely ignores the
fundamental physics behind the equation of state. However, with the purpose to
make connection with the current literature I will reformulate the results
above in terms of the slow-roll potentials.

To determine the potential $V\left(  \varphi\right)  \simeq\varepsilon$ we
have to express $N$ in terms of the scalar field $\varphi.$ With this purpose
we write%
\begin{equation}
\frac{d\varphi}{dN}=\frac{\dot{\varphi}}{-H}=\sqrt{\frac{3}{8\pi}\left(
1+w\right)  },\label{11}%
\end{equation}
where we have taken into account that $\varepsilon+p=\dot{\varphi}^{2}$ and
$H^{2}=8\pi\varepsilon/3$ (in the Planck units). Substituting here (\ref{19a})
and integrating the resulting equation we find than for $\alpha=1,$
\begin{equation}
\varphi\simeq\left\{
\begin{array}
[c]{c}%
\sqrt{N},\text{ \ \ \ \ \ \ \ \ \ \ \ \ \ \ \ \ \ \ \ \ }N_{m}-N>N\\
\varphi_{m}-\sqrt{N_{m}-N},\text{ \ \ \ \ }N_{m}-N<N\text{\ }%
\end{array}
\right.  \label{12}%
\end{equation}
Taking this into account in expression $\left(  \ref{20a}\right)  $ we find
that the even in $\varphi$ potential which well approximates the required
behavior of the equation of state is
\begin{equation}
V\left(  \varphi\right)  \simeq\frac{\varphi_{m}^{a}\left(  \varphi
/\varphi_{m}\right)  ^{2}}{\left(  \varphi_{m}^{2}-\varphi^{2}\right)  ^{a}%
},\label{26}%
\end{equation}
where $a=2/\gamma$ and can be for instance $3,4...$ In fact, for positive
$\varphi$%
\begin{equation}
1+w\simeq\left(  \frac{V_{,\varphi}}{V}\right)  ^{2}\simeq\frac{1}{\varphi
^{2}}\left(  \frac{\varphi_{m}}{\varphi_{m}-\varphi}\right)  ^{2},\label{27}%
\end{equation}
and inflation takes place for $1<\varphi<\varphi_{m}-1.$ The spectrum of
perturbations is given by%
\begin{equation}
\Phi\simeq\frac{V^{3/2}}{V_{,\varphi}}\simeq\frac{\left(  \varphi/\varphi
_{m}\right)  ^{2}}{\left(  \varphi_{m}-\varphi\right)  ^{\frac{a}{2}-1}%
}\label{28}%
\end{equation}
and $\Phi$ becomes of order unity only at Planck scale when $\varphi
_{m}-\varphi\simeq1.$ To get the required amplitude of perturbations in
galactic scales we have to take%
\begin{equation}
\varphi_{m}\simeq10^{\frac{12}{a+2}}\label{29}%
\end{equation}
For $a=4$ we have $\varphi_{m}\simeq10^{2}$ and correspondingly the total
number of e-folds on inflation is $N_{m}\simeq\varphi_{m}^{2}\simeq10^{4}$. As
we have noticed above the potential $\left(  \ref{26}\right)  $ predicts too
much gravitational waves. 

More realistic potential, which is in perfect agreement with the observations,
is%
\begin{equation}
V\left(  \varphi\right)  \simeq\frac{\left[  1-\exp\left(  -\varphi\right)
\right]  ^{2}}{\left(  \varphi_{m}-\varphi\right)  ^{a}}.\label{30}%
\end{equation}
For this potential%
\begin{equation}
1+w\simeq\left(  \frac{V_{,\varphi}}{V}\right)  ^{2}\simeq\left(  \frac
{a}{\varphi_{m}-\varphi}+2e^{-\varphi}\right)  ^{2},\label{31}%
\end{equation}
and
\begin{equation}
\Phi\simeq\frac{V^{3/2}}{V_{,\varphi}}\simeq\frac{1}{\left(  \varphi
_{m}-\varphi\right)  ^{\frac{a}{2}-1}\left(  a+2e^{-\varphi}\varphi
_{m}\right)  }\label{32}%
\end{equation}
for $1<\varphi<\varphi_{m}.$ The value of $\varphi_{m}$ is entirely fixed by
the required amplitude of perturbations and is equal to $\varphi_{m}%
\simeq10^{\frac{12}{a}}.$ The possible values for $a$ are $4$ and $6.$ For
$\varphi<\ln\varphi_{m}$ we obtain from $\left(  \ref{11}\right)  $ $N\simeq
e^{\varphi},$ while for $\varphi>\ln\varphi_{m}$ one has $N_{m}-N\simeq\left(
\varphi_{m}-\varphi\right)  ^{2}.$ Therefore this model is similar to the
model $\left(  \ref{19a}\right)  $ with $\alpha=2$ and the potential $\left(
\ref{30}\right)  $ is of type which is the best from the point of view of the
CMB observations and at the same time does not lead to the problems of initial
conditions and multiverse, thus recovering the \textquotedblleft classic
inflationary paradigm\textquotedblright.

\section{Discussion}

The Planck measurements have unambiguously confirmed the main predictions of
the theory of quantum origin of the universe structure. Namely, the adiabatic
nature and the Gaussian origin of primordial perturbations were established
beyond any reasonable doubt. Even more amazing, more nontrivial infrared
logarithmic tilt of the spectrum, first predicted in \cite{M2}, was discovered
at 6 sigma confidence level. The simplest way to amplify the quantum
fluctuations is provided by the stage of inflation. Although nobody doubt the
quantum origin of the primordial fluctuations, there are still claims in the
literature that basically the same mechanism of amplification of quantum
fluctuations can work also either in a bouncing universe on the stage of super
slow contraction \cite{ST} or in conformal rolling scenario \cite{rubakov}.
The generated spectra in the alternative theories are not the predictions of
the theory, but rather postdictions which are constructed to be in agreement
with observations. Nevertheless, this is not enough to rule out these
possibilities at the level of a "theorem". Thus, at the moment the only
robustly established experimental fact is the quantum origin of the universe
structure with a little uncertainty left for the mechanism of amplification of
quantum fluctuations. To firmly establish that namely inflation has provided
us this mechanism one has to find the primordial gravitational waves the lower
bound on which for the spectral index $n_{s}=0.96$ corresponds to $r$ about
$0.003.$ According to \cite{ILS1, ILS2} one of the main motivations for
looking the alternatives to inflation is the failure of predictability of so
called "postmodern inflationary paradigm". Paradoxically this trouble seems to
be due to the same successful quantum fluctuations with the red-tilted
spectrum which lead to the galaxies. On one hand the quantum fluctuations
explain the observed large scale structure of the universe, but on the other
hand they are also responsible for the selfreproduction and produce eternal
inflating multiverse where "anything can happen and will happen an infinite
number of times" \cite{GKN}. In this paper I have shown how this problem can
be avoided. Using the effective description of inflation I have found nearly
unambiguous extension of inflation which avoids the selfreproduction. What is
yet missing in this description is a justification of the model from the point
of view of some fundamental theory. However, under circumstances when only
effective description of inflation is needed to explain the observations and
there are no even slightest experimental hints how the fundamental theory
should look like at very high energies such an approach looks as the most
plausible. Moreover, it can provide us with hints about fundamental theory,
which can avoid even metaphysical problems.

\textbf{Acknowledgements} This work was supported by TRR 33 \textquotedblleft
The Dark Universe\textquotedblright\ and the Cluster of Excellence EXC 153
\textquotedblleft Origin and Structure of the Universe\textquotedblright.

\end{document}